\documentclass[aps,prl,nobalancelastpage,superscriptaddress,twocolumn,longbibliography,nobibnotes,nofootinbib]{revtex4-2}

\usepackage[utf8]{inputenc}
\usepackage[american,british]{babel}
\usepackage[T1]{fontenc}
\usepackage[pdftex]{graphicx}  
\usepackage{xcolor}
\usepackage{dcolumn}
\usepackage{physics}
\usepackage{braket}
\usepackage{bm}
\usepackage{amsmath,amsthm,amssymb}
\usepackage{color}
\usepackage{verbatim}
\usepackage[normalem]{ulem}
\usepackage{hyperref}
\hypersetup{
 colorlinks=true,
 linkcolor=blue,
 anchorcolor = blue,
 citecolor = blue,
 filecolor = blue,
 urlcolor = blue
}

\def \be {\begin{equation}} 
\def \ee {\end{equation}} 
\def \l {\left(} 
\def \r {\right)} 
\def \la {\langle} 
\def \ra {\rangle}  

\date{}

\newcommand{\saclay}{Universit\'e Paris-Saclay, CNRS, LPTMS, 91405, Orsay, France.}

\begin{document}

\title{
Emergent integrable dynamics in a non-integrable Rydberg-atom chain
}
\date{\today}

\author{Gianluca~Morettini}
\affiliation{\saclay}

\author{Luca Capizzi}
\affiliation{\saclay}

\author{Leonardo~Mazza}
\affiliation{\saclay}
\affiliation{Institut Universitaire de France, 75005, Paris, France}

\author{Maurizio Fagotti}
\affiliation{\saclay}

\begin{abstract}
We show that integrable and non-integrable dynamics can coexist in the same Rydberg-atom chain, depending on the initial state in which the system is prepared.
In the setting we consider, Rydberg atoms mainly experience an effective dipolar interaction and, within the nearest-neighbor approximation, their dynamics can be mapped onto an effective integrable Fermi gas with ballistic transport.
The inclusion of longer-range couplings, however, is essential for the theory to be predictive; those terms break the conservation laws associated with integrability, enabling, in particular, the emergence of genuine diffusive transport.
We study the dynamics
generated by a bipartition protocol and reveal a sharp qualitative
change in behavior as the particle density is varied, suggesting the
possibility of accessing weaker and stronger integrability breaking dynamics in the same Hamiltonian depending on the relevance of local interactions.
We propose a theoretical mechanism that accounts for the differences.
Our findings provide clear and concrete evidence that integrability breaking is not solely a property of the Hamiltonian and of the magnitude of the couplings that break integrability, but also of the state in which the system is prepared.
\end{abstract}

\maketitle 

\paragraph{Introduction ---}

Understanding thermalization in generic isolated systems can be difficult because the boundary between integrable and non-integrable dynamics is not sharp.
This is especially evident when the interactions that break integrability are associated with coupling constants that are small relative to the accessible timescales.
In classical mechanics, the Kolmogorov–Arnold–Moser (KAM) theorem~\cite{k-54,n-71,p-93} provides a precise characterization of weak integrability breaking and of its timescales;
% the typical timescales on which its effects become visible;
in quantum theory, the existence of a KAM-type stability has not yet been pinpointed.
The situation is particularly complex in one-dimensional settings, classical or quantum, both because of the stringent kinematic constraints,
and because integrable points may be hidden throughout the parameter space~\cite{Fermi-55}.
 
Interest in integrable quantum systems has been profoundly stimulated by recent experimental progress~\cite{Kinoshita-06,Langen2015Science,sbdd-19,Malvania2021,Wei2022,Dubois_2024,Schuettelkopf2025}; this has motivated the extension to integrable systems of both statistical mechanics~\cite{Essler_2016, Vidmar_2016} and hydrodynamics~\cite{Bertini-16, Castro-16},
% leading to the development of generalized Gibbs ensembles and generalized hydrodynamics (GHD)~\cite{Bertini-16, Castro-16},
tested in high-precision experiments~\cite{Amerongen_2008, sbdd-19}. 
However, the intermediate times accessible in experiments often coincide with the regime in which non-integrable dynamics blend with integrable ones.
On the one hand, there is extensive evidence of integrable dynamics in realistic systems, beginning with the landmark ``quantum Newton’s cradle'' experiment, which revealed the remarkable persistence of near-integrable behavior in quasi-one-dimensional Bose gases~\cite{Kinoshita-06}. On the other hand, a large body of theoretical work has predicted and described the late-time emergence of diffusion induced by integrability-breaking interactions~\cite{Bastianello-21}.
By contrast, in this renewed effort to address the long-standing problem of thermalization in quantum many-body systems, comparatively little attention has been paid to the role of the initial state in the relaxation process. 

In this letter we identify a quantum many-body setup in which integrable and non-integrable dynamics coexist depending on the initial conditions. 
We consider a dipolar XY model, which provides a satisfactory description of the dynamics in an experimentally realized chain of Rydberg atoms~\cite{Chen-26} and provide compelling evidence that this model is not integrable, although it is close to an integrable one. 
We consider bipartitions obtained joining together ground states at different magnetizations:
we identify the configurations whose dynamics exhibit a ballistic behavior that is well captured by effective free-fermion models;
more generally, however, this simple picture breaks down. 
This behavior is probed by studying both the expectation values and fluctuations of observables that can be detected experimentally~\cite{Schauss-15, Bernien-17,Schymik_2020,Browaeys-20,Ebadi-21, Semeghini-21,Rosenberg_2024,2025_Chen,Emperauger-25,Chen-26}. 
This result highlights from a hydrodynamic perspective the pseudo-local nature of the conservation laws that characterize the dynamics at mesoscopic space- and time-scales.
From a quantum-chaos perspective this work exhibits scenarios that may be viewed as the quantum counterpart of the weak-integrability-breaking regime at the heart of the KAM theorem in classical mechanics.
From a practical perspective, it provides an experimental proposal to access such dynamics.

\paragraph{The setup ---}

Recent progress in Rydberg atomic-array platforms~\cite{Nogrette_2014, Schauss-15, Barredo2016, Bernien-17, Browaeys-20, Schymik_2020, Ebadi-21, Semeghini-21,Bornet2023,Emperauger-25} has enabled the realization of experimental setups that are well-described by the one-dimensional dipolar XY model~\cite{Chen-23,Chen-26}
\begin{equation}\label{eq:XY_ham}
H = -J\sum_{i<j}\frac{1}{|i-j|^3}(\sigma^+_i\sigma^-_j+\sigma^+_j\sigma^-_i).
\end{equation}
When Hamiltonian~\eqref{eq:XY_ham} is approximated by retaining only the nearest-neighbor terms ($j=i+1$), the resulting model can be mapped onto free fermions 
\begin{equation}\label{eq:H0}
H_0 = -J\sum_{j}[\psi^\dagger_j\psi_{j+1} + h.c.]
\end{equation}
via the Jordan-Wigner transformation~\cite{jw-28}, $\sigma^{-}_j = \exp\l i\pi \sum_{j'<j} \psi^\dagger_{j'} \psi_{j'}\r \psi_j$, where $\psi_j$ is a fermionic annihilation operator at site $j$; 
$H_0$ is therefore integrable.

Integrability is broken by interactions between more distant spins, which, despite the smallness of the corresponding coupling constants, leave measurable signatures in the dynamics of magnetization fluctuations, even in small arrays of Rydberg atoms~\cite{Chen-26}.
This was traced back to the impossibility of deforming a subset of the conservation laws in such a way as to allow for a smooth modification of the integrable dynamics on the intermediate timescale of the experiment.
The possibility of distinguishing genuinely integrable dynamics from those governed by an incomplete set of quasi-conservation laws using a  quantum simulator opens the exciting perspective of searching for initial states that exhibit emergent integrable behavior.

We simplify Hamiltonian~\eqref{eq:XY_ham} by truncating the sum at the next-to-nearest-neighbor term, since the latter captures the main integrability-breaking effects on the timescale of interest. We thus write the Hamiltonian as $H = H_0 + \eta V$, with $\eta = 1/8$, separating the integrable part from the perturbation, which we write in the more transparent fermionic representation
\begin{equation}\label{eq:pert}
V = -J\sum_{j}[\psi_j^\dagger\psi_{j+2}-2\psi_j^\dagger\psi^\dagger_{j+1}\psi_{j+1}\psi_{j+2} + h.c.].
\end{equation}
Two contributions appear in Eq.~\eqref{eq:pert}: the first, quadratic, corrects the dispersion relation of the free particles, while the second is quartic and introduces interactions that break integrability; see the End Matter (EM) for details, including our study of the level-spacing statistics.

\paragraph{Kinematics of two-particle scattering ---}

In order to study the effect of the perturbation $V$ reported in Eq.~\eqref{eq:pert}, which is the leading source of inter-particle scattering in Hamiltonian~\eqref{eq:XY_ham}, we first focus on the two-particle problem.
The unperturbed energy-momentum dispersion relation of the quadratic hamiltonian $H_0$ reads $\varepsilon(k) = -2J\cos(k)$.
We will deliberately ignore the renormalization of the single-particle dispersion relation associated with the perturbation in Eq.~\eqref{eq:pert} because we want to emphasize the aspects that modify the dynamics qualitatively rather than merely quantitatively. 

Within this approximation, we consider a pair of incoming quasiparticles with momenta $(k_1, k_2)$ and let them scatter, denoting the momenta of the outgoing quasiparticles by $(k_1', k_2')$. Conservation of energy and momentum (modulo $2\pi$, because of the lattice) requires
\begin{equation}\label{eq:constraints}
\begin{cases}
\varepsilon(k_1) + \varepsilon(k_2) = \varepsilon(k_1') + \varepsilon(k_2'),\\
k_1 + k_2 \equiv k_1' + k_2' \mod 2 \pi.
\end{cases}
\end{equation}
Equation~\eqref{eq:constraints} strongly constrains the kinematics of the scattering problem, regardless of the specific form of the interaction. While a trivial solution, in which the outgoing momenta coincide with the incoming ones, is always allowed, additional nontrivial solutions arise only for pairs $(k_1, k_2)$ with total momentum $k_1 + k_2 = \pi$. Specifically, any such pair can scatter into any other pair with total momentum $\pi$, since their total energy is always equal to $0$.

It is interesting to observe that only half of the shift-invariant extensive and local conservation laws of Hamiltonian~\eqref{eq:H0} are conserved by these scattering processes, and are henceforth dubbed \textit{robust}, we call the remaining ones \textit{fragile}~\cite{Burgarth2021}.
To see this, we introduce the  fermionic operators $c_k^{(\dagger)}$ in Fourier space, and recall that the conservation laws read $Q_{1,n} = \int \frac{dk}{2 \pi} \cos(nk) c_k^\dagger c_k $ and $Q_{2,m} = \int \frac{dk}{2 \pi} \sin(mk) c_k^\dagger c_k$, with $n,m \in \mathbb N$~\cite{Grabowski_1995}.
It is an easy calculation to show that the allowed scattering processes do not alter the values of the conservation laws $Q_{1,n}$ with odd $n$, and $Q_{2,m}$ with even $m$, namely those whose associated trigonometric function is zero for $k = \pi/2$.

This scenario is explicitly ruled out in systems with either Galilean invariance ($\varepsilon(k)=k^2/{2m}$) or relativistic invariance ($\varepsilon(k)=\sqrt{k^2+m^2}$): there, the kinematical constraints are so restrictive that the only solution to \eqref{eq:constraints} is the trivial one: the additional scattering channel is a lattice effect. Finally, we remark that this analysis is intended to hold only within the perturbative regime of small $\eta$. For finite values of $\eta$, the single-particle dispersion $\varepsilon(k)$ undergoes nontrivial renormalization and the conclusions derived from the kinematic constraints in Eq.~\eqref{eq:constraints} should be modified accordingly.

\paragraph{Dynamics under the action of a weak perturbation ---}

To scale up the results for the two-particle scattering to the many-body setting, we investigate the dynamics of a stationary state of $H_0$ under $H_0 + \eta V$.
Let us consider for instance the following initial state
\begin{equation}\label{eq:n_NESS}
n(k) = \begin{cases} 1 \quad k \in [0,k_\text{F}],\\ 0 \ \text{ otherwise},
\end{cases}
\end{equation}
describing a Non-Equilibrium Steady State (NESS) of the unperturbed model.
The left-moving modes $k \in [-\pi,0]$ of such a state are unoccupied and a finite and positive current, given by $j = \int \frac{dk}{2\pi} \varepsilon'(k)n(k) = \frac{J}{\pi}(1-\cos(k_F))$, is carried by the right-moving quasiparticles.

When $k_{\text{F}} > \pi/2$, pairs of particles with positive momenta can undergo backscattering, thereby populating modes with negative momenta. One such process degrades the current flowing in the NESS, and is responsible for the disappearance of a ballistic transport phenomenology. Given that the model $H_0 + \eta V$ is not integrable, this result is not surprising.

Conversely, for $k_{\text{F}} < \pi/2$,  no pairs of particles with total momentum $\pi$ are present: the NESS is stable and, at leading order, no scattering process can take place: current degradation appears on timescales qualitatively longer than those of the previous case.
Remarkably, this is not associated with an initial state with zero fermionic density, as whenever $k_F$ is finite there is a finite density of particles in the thermodynamic limit.
At fixed timescales, thus, the same Hamiltonian can produce qualitatively different dynamics which are sensitive to the perturbation at different orders.

\paragraph{The bipartition protocol ---}

To investigate this scenario in an actual quantum many-body system with a controlled numerical simulation, we consider a bipartition protocol with an initial state that can be prepared experimentally in modern Rydberg platforms (see Ref.~\cite{Emperauger-25} for details).
We consider the Hamiltonian $H_0 + \eta V$ restricted to the left half-chain (whose length is $L/2$), with open boundary conditions, and we prepare its ground state in the subspace with a fixed number of fermions $N$.
Similarly, we prepare the right half-chain in the vacuum without fermions (i.e., the ground state with $N=0$ fermions). Finally, we let the system evolve with the total Hamiltonian, acting on the chain of length $L$ with open boundary conditions.

Let us comment on the initial state in the left half of the chain.
For $\eta=0$, this state corresponds to a Fermi sea, where the occupied modes are $k\in [-k_{\text{F}},k_{\text{F}}]$: the correspondence between the Fermi momentum $k_{\text{F}}$ and the particle density is given by
\be\label{eq:k_F}
\frac{k_{\text{F}}}{\pi} = \frac{N}{L/2}.
\ee
For small $\eta$, although interactions are present, the salient features of the ground state are similar to those of the free case:
the Luttinger theorem~\cite{Luttinger-60} shows rigorously this is the case for short-range interactions, and recent experiments suggest the same for Rydberg systems with $1/r^3$ interactions~\cite{Emperauger-25}. 
For this reason, we will employ Eq.~\eqref{eq:k_F}, strictly exact in the absence of interactions, as a definition of $k_{\text{F}}$ for the weakly interacting ground state.

\begin{figure}[t]
\includegraphics[width=0.7\columnwidth]{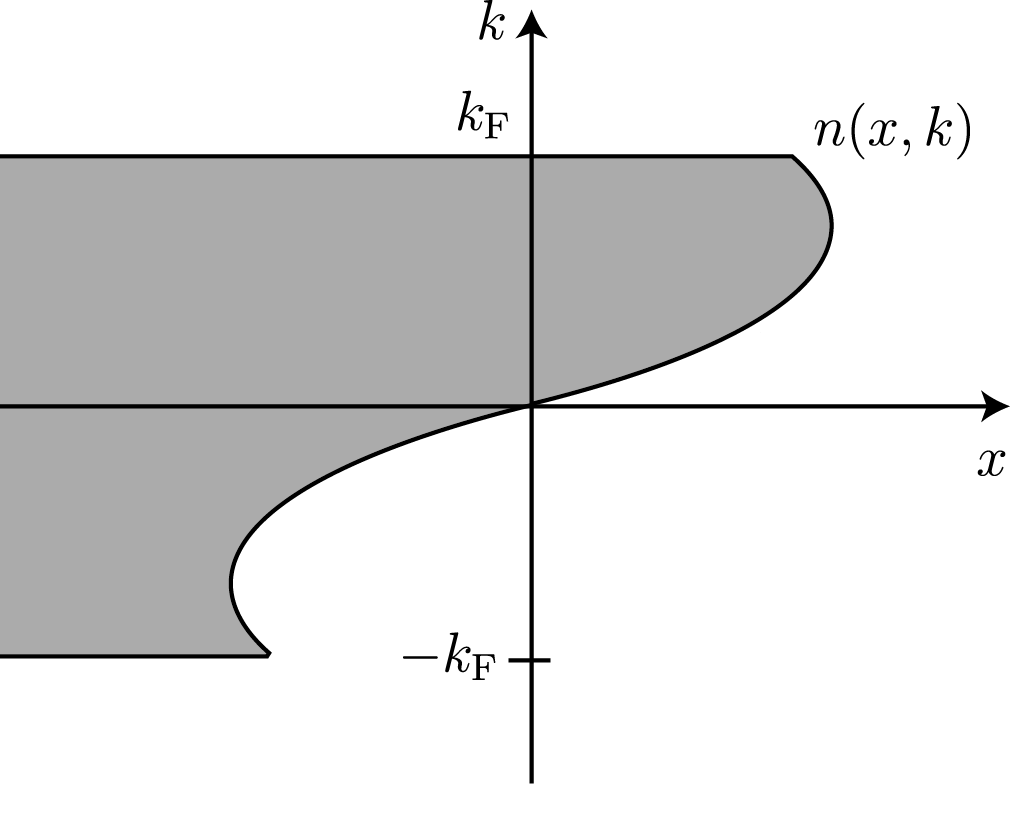}
 \caption{The local occupation $n(x,k)$ as a function of the position $x$ and the momentum $k$ for $t > 0$, evolved from the initial condition \eqref{eq:n_initial} with the free-fermionic hydrodynamic equation \eqref{eq:free_GHD}. The function $n(x,k)$ equals $1$ within the gray regions and $0$ otherwise. }
 \label{fig:Fermi_sea}
\end{figure}

Before commenting on the interacting case, it is worth reminding the properties of the dynamics in the absence of perturbations. When $\eta =0$, the large-scale dynamics (at the so-called Euler scale) can be described via GHD equations for the local occupation function at position $x$, denoted by $n(x,k)$, and namely~\cite{Antal_1999,Fagotti-17,Fagotti-20}
\begin{equation}\label{eq:free_GHD}
\partial_t n(x,k) + \partial_x [\varepsilon'(k)n(x,k)]=0,
\end{equation}
with initial conditions
\begin{equation}\label{eq:n_initial}
n(x,k) = \begin{cases}1, \quad x<0 \text{ and } k\in [-k_{\text{F}},k_{\text{F}}],\\ 0, \quad \text{ otherwise}.\end{cases}
\end{equation}
We sketch the corresponding solution in Fig.~\ref{fig:Fermi_sea}: in particular, the associated \textit{non-equilibrium steady state} (NESS), obtained in the limit of large $t$ for fixed $x$ is described by the occupation function in Eq.~\eqref{eq:n_NESS}.
According to this theory, the density profile $\langle \psi_j^\dagger \psi_j \rangle (t)$ should display a ballistic rescaling and only depend on $x/t$.
Previous studies have shown that in non-interacting systems, $\eta=0$, the variance of the half-chain magnetization
$
Q_A = \sum_{j \in A} \psi^\dagger_j\psi_j
$, 
with $A = [-L/2,0]$, denoted by $\la Q^2_A\ra_c$, grows logarithmically in time $\sim \log t$; a similar behaviour is expected for the half-chain entanglement entropy~\cite{Eisler-2014,Dubail_2017_SciPost,Scopa_2021,Scopa-22,Scopa_2022_SciPost,Ares_2022,Scopa-23}.

When $\eta \neq 0$, the dynamics of the system is not exactly solvable. 
However, according to the analysis presented so far, for $k_{\text{F}} \lesssim  \pi/2$ and on short and intermediate length- and time-scales, the dynamics should be perturbatively close to those of non-interacting fermions we just described. 
Indeed, the 2-particle scattering cannot play a role since no pairs of particles with total momentum $\pi$ are present. 
By contrast, for $k_\text{F} \gtrsim  \pi/2$ the 2-particle scattering processes are leading and the dynamics change qualitatively. 

We stress that higher-order scattering events can occur in the full many-body dynamics, but they are related to higher-order effects in perturbation theory, and thus are expected to have a role only at much later times. 
For this reason they have not been taken into account in the main discussion (see EM for details): on short and intermediate scales they should not play a significant role. To verify this expectation,
we study the dynamics of these quantities in this setup for $\eta \geq 0$ with an algorithm based on the MPS Itensors library~\cite{ITensor, ITensor-r0.3}, which allows to study one-dimensional quantum many-body dynamics in a controllable way~\cite{SCHOLLWOCK201196}.

\begin{figure}[t]
\includegraphics[width=\columnwidth]{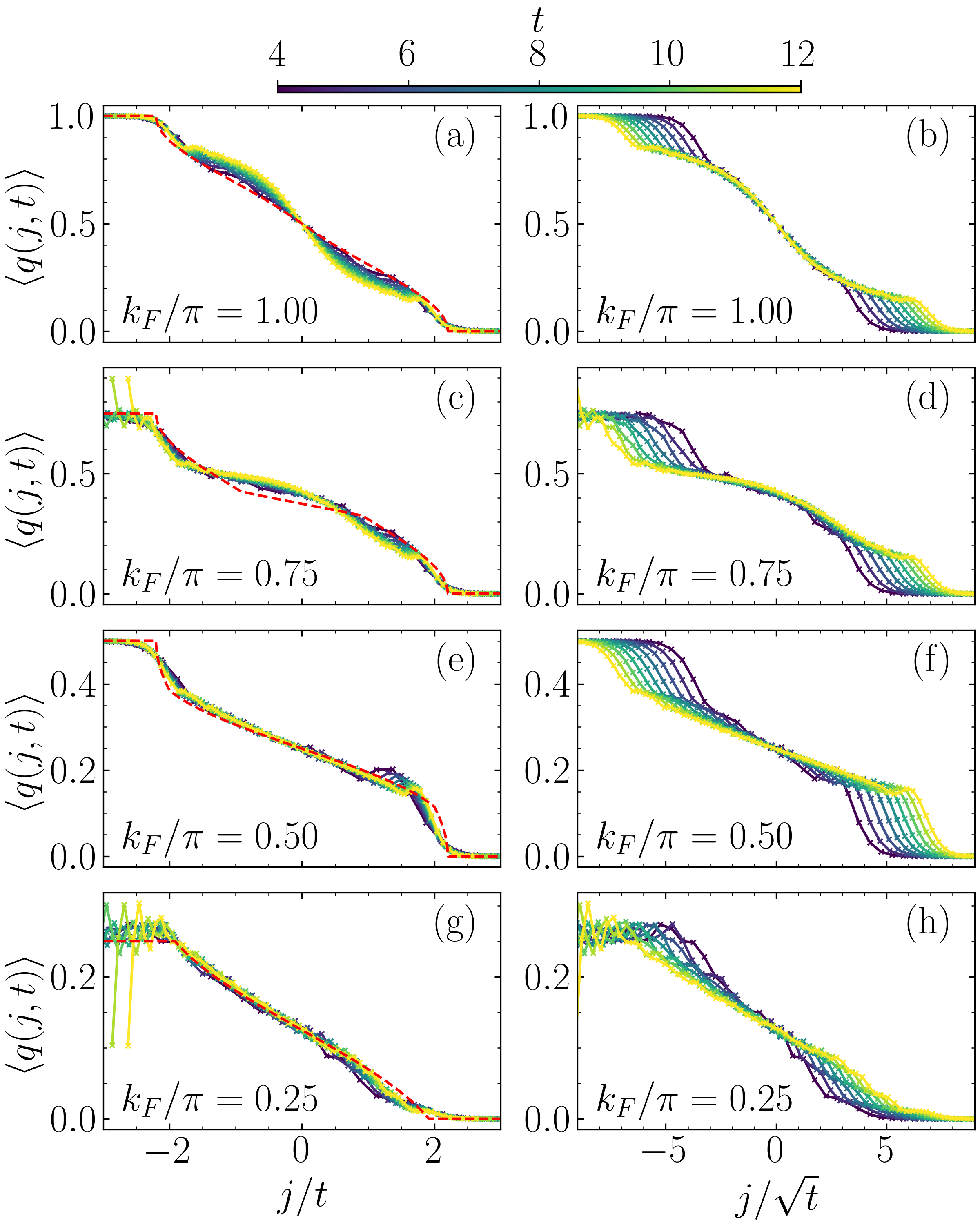}
 \caption{Spatial profile of the charge density for different values of $k_{F}$. Left panels: Plot as a function of $j/t$ (ballistic rescaling); we compare the data with a free fermionic approximation (dashed red line). Right panels:  Plot as a function of $j/\sqrt{t}$ (diffusive rescaling). The parameters are $[J,L,\eta] = [1, 64,1/8]$.
 }
 \label{fig:Profile_Z}
\end{figure}

\paragraph{Numerical study: Magnetization profile---}
We begin by studying the space-time structure of the dynamics of local observables. Specifically, we consider the fermionic density $q(j) := \psi^\dagger_j\psi_j$ and study its expectation value as a function of time, denoted by $\la q(j,t)\ra$; in Fig.~\ref{fig:Profile_Z} we plot it as a function of both $j/t$ and $j/\sqrt{t}$ for $t\leq 12$. 
The panels, which refer to different values of $k_F$, support that the fermionic density has a ballistic scaling for $k_F/\pi = 0.25$ and $0.5$, whereas a diffusive scaling is more suitable for $k_F = 0.75$ and $1$. 
In fact, a careful analysis of panels (\textit{a}) and (\textit{c}) shows that the edges of the light-cone profile still obey a ballistic rescaling. 
This twofold nature of the magnetization density is intriguing and deserves a more in-depth study.

To further stress the link between ballistic behaviour for $k_F/\pi \leq 0.5 $ and an asymptotic integrable behaviour,
we compare our numerical data  with an approximated hydrodynamic free-fermionic prediction, obtained from Eq.~\eqref{eq:free_GHD} with the simplified dispersion relation $\varepsilon(k) = -2J (\cos k +\eta \cos (2k))$, which neglects the quartic term in Eq.~\eqref{eq:pert}. 
The result, plotted in red, describes in a satisfactory way the data for $k_F \leq 0.5$.
We mention that the small discrepancies are well captured by an asymptotic TBA equation, which however deserves a separate discussion.

\begin{figure}[t]
\includegraphics[width=\columnwidth]{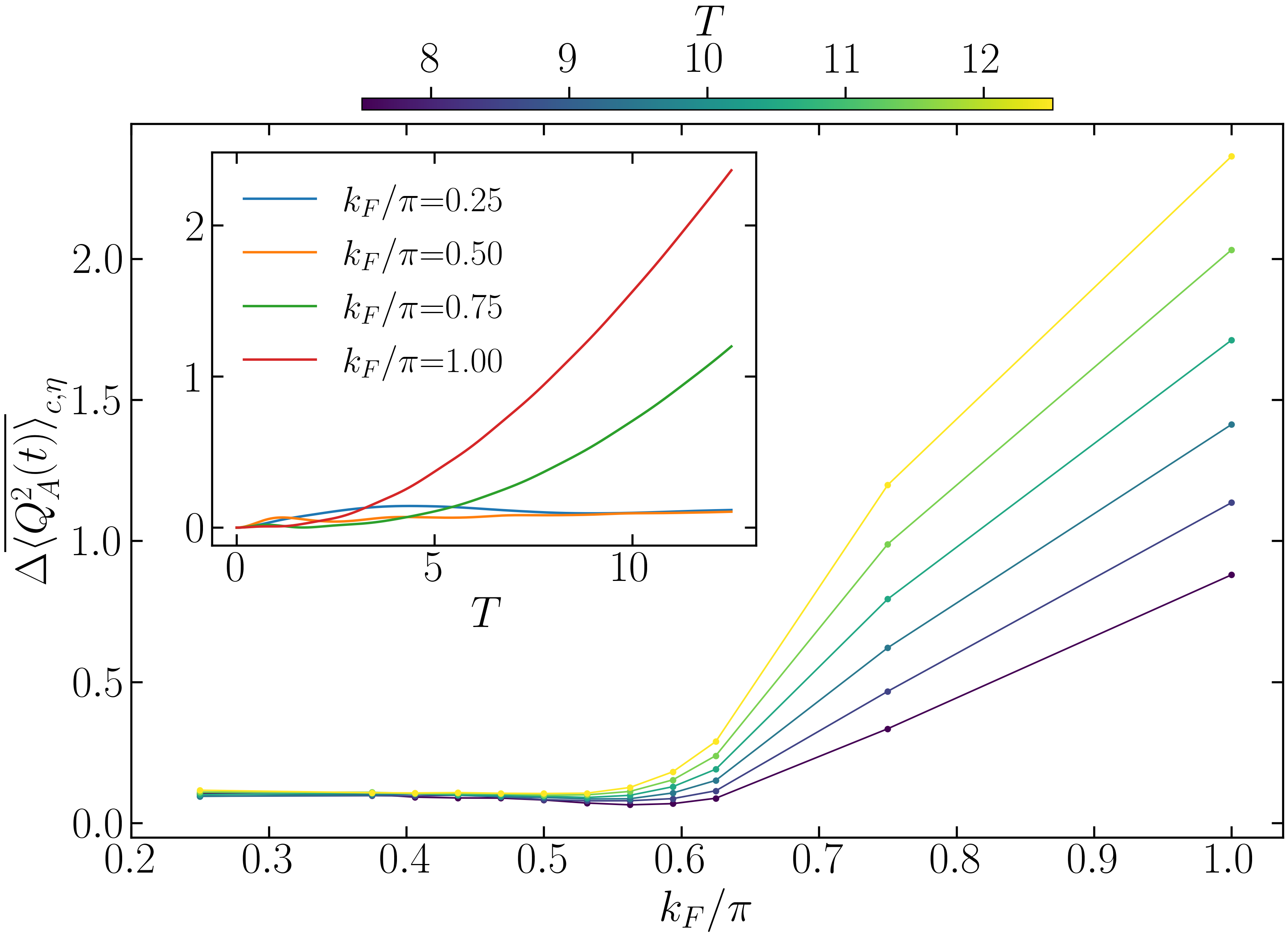}
 \caption{Difference between the magnetization variance in the presence of the perturbation (at $\eta= 1/8$) and in its absence, averaged over the time window $t \in [0,T]$ and plotted as a function of $k_{\text{F}}$. An inset show the same quantity as a function of $T$. The parameters are $[J,L] = [1, 64]$. 
 }
 \label{fig:Transition}
\end{figure}

\paragraph{Numerical study: Variance of the half-chain magnetization ---}
We analyze the growth of the half-chain magnetization, which was recently studied in a Rydberg-atom quantum simulator to identify the presence of broken conservation laws~\cite{Chen-26}. 
We average over a time window $t \in [0,T]$,
\begin{equation}\label{eq:av_var}
\overline{\la Q^2_A(t)\ra}_{c,\eta} := \frac{1}{T}\int^T_0 dt \la Q^2_A(t)\ra_{c,\eta};
\end{equation}
and subtract from the data obtained for $\eta=1/8$ (relevant for Rydberg-atom quantum simulators) its value at $\eta =0$: 
\begin{equation}
    \Delta \overline{\la Q^2_A(t)\ra}_{c,\eta} := 
    \overline{\la Q^2_A(t)\ra}_{c,\eta} -
    \overline{\la Q^2_A(t)\ra}_{c,\eta=0}.
\end{equation}
This definition helps highlighting the behaviour of $\overline{\la Q^2_A(t)\ra}_{c,\eta}$ that is only due to the specific effect of the integrability-breaking perturbation.

Our results are shown in
Fig.~\ref{fig:Transition}.
The inset displays the behaviour
of $\Delta \overline{\la Q^2_A(t)\ra}_{c,\eta}$ as a function of time for the four values of $k_F$ considered in Fig.~\ref{fig:Profile_Z}. Two different qualitative behaviours appear: for $k_F/\pi \leq 0.5 $ it is approximately constant, signaling a growth $\sim \log(t)$ as in the integrable case; for $k_F/\pi \geq 0.75$ a sizeable growth appears, which is compatible with a power-law behavior.
In the main panel we present $\Delta \overline{\la Q^2_A(t)\ra}_{c,\eta}$ as a function of $k_{\text{F}}$ for different values of the time $T$. Our data are consistent with a significant increase of the variance for $k_{\text{F}} \gtrsim \pi/2$, whereas for $k_{\text{F}} \lesssim \pi/2$ the variance remains close to its initial value.
The change of behaviour as a function of $k_F$ is sharp and abrupt, marking two regimes that are qualitatively different: for this reason, the integrable behaviour appearing for $k_F  \lesssim \pi/2$ should not be thought of as an extremal behavior of the interacting situation, but rather as a qualitatively different one. On the time-scales accessible to our numerics, the setup is indistinguishable from an effective integrable model.
Similar results are presented in the EM for the growth of the bipartition entanglement entropy.

Finally, although we presented numerics only for a bipartition protocol based on joining two ground states, the theoretical interpretation based on the 2-particle scattering mechanism makes clear that any bipartition protocol obtained joining together states with occupied momenta such that $|k| \lesssim \pi/2$ display a dynamics that is qualitatively equivalent to that of free-fermion chains. This extends our result to a wide class of initial states scaling exponentially in the system-size length.

\paragraph{Conclusions ---}

As we stressed in the introduction, experiments such as the "quantum Newton cradle" suggest that the notion of integrability applied to experimental setups should be intended as \textit{a theoretical tool with a descriptive power on given time- and length-scales}.
% What we just discussed suggests the remarkable fact that i
We showed that in a setup where the conservation laws of an integrable model have been weakly broken there are initial states that are sensitive to integrability breaking at different levels and \textit{sometimes, at short and intermediate times, behave as if the model were integrable}.

We studied the dynamics of the magnetization and of its variance in a non-integrable Rydberg lattice system initialized according to the bipartition protocol. 
Guided by an intuitive theory based on 2-particle scattering, we presented controlled numerical simulations demonstrating a sharp change from ballistic to diffusive transport at finite magnetization density.
At fixed Hamiltonian, hence, different initial states can be sensitive to integrability breaking at different perturbative orders.
Our model is routinely employed to describe state-of-the-art Rydberg-atom platforms and the observables we considered are accessible~\cite{Chen-26}: the experimental observation of these effects in the near future is possible.

Our results shed new light on the long-standing theoretical problem of weak-integrability breaking in one-dimensional quantum systems and suggest that a non-integrable model with a well-defined Wigner-Dyson level-spacing statistics may support a thoroughly integrable dynamics on short and intermediate length- and time-scales, if prepared in an appropriate initial state.
The coexistence of integrable and non-integrable dynamics in the same quantum spin chain resonates with early studies on classical chaos.
% Our findings provide clear and concrete evidence that integrability breaking is not solely related to the magnitude of the couplings that break integrability, but also to the state in which the system is prepared.
Moreover, our findings go beyond the conventional picture in which weak integrability breaking merely gives rise to prethermal behavior.

Beyond identifying the best conditions to perform this study with current quantum simulators,
several theoretical questions remain open. Specifically, since for $k_F \leq \pi/2$ the system is qualitatively well-described by  free fermionic dynamics, it is reasonable to assume that a self-consistent mean-field approximation 
(in the same spirit of Ref. \cite{Robertson-23}) 
can be quantitatively accurate.
Once such theory is prepared,
one can speculate to  incorporate the effects of the perturbation order by order, obtaining from microscopic principles a set of hydrodynamic equations that are descriptive for all values of $k_F$. We plan to come back to this problem.

\paragraph{Acknowledgements ---}
We acknowledge discussions on closely related theoretical projects with F.~Ferro, F.~Kahlert and A.~Marché and with the experimental team at Institut d'Optique Graduate School (C.~Chen, G.~Bornet, G.~Emperauger, Th.~Lahaye and A.~Browaeys).
We also thank A.~Bastianello, J.~De Nardis, F.~Essler, S.~Murciano, S.~Scopa, F.M.~Surace, S.~Thomsovic and D.~Ullmo for insightful comments on these results.
We acknowledge support from the ANR project LOQUST ANR-23-CE47-0006-02 (LC and LM) and by the PEPR Dyn-1D ANR-23-
PETQ-0001 (LM).
This work is part of HQI (www.hqi.fr) initiative and is supported by France 2030 under the French National Research Agency grant number ANR-22-PNCQ-0002 (LM). GM is supported by the QuanTEdu-France program (State grant part of France 2030, grant ANR-22-CMAS-0001).

\begin{center}
\begin{large}
\textbf{End Matter}
\end{large}
\end{center}

\paragraph{Non-integrability test---}

We discuss the level-spacing statistics of $H_0 + \eta V$ for different values of $\eta$. In particular, we consider the eigenvalues of the Hamiltonian in the middle of the spectrum, ordered as
$E_{N/3} \leq E_{N/3+1} \leq \ldots \leq E_{2N/3-1} \leq E_{2N/3}$.
We define the normalized level spacing as $s_i = (E_{i+1} - E_i)/\delta$, where
$\delta = (E_{2N/3} - E_{N/3})/(N/3)$ is the mean level spacing.
From $s_i$, one can also define the ratio
$r_i = \min(s_i, s_{i-1})/\max(s_i, s_{i-1})$,
which is independent of $\delta$, together with its statistical average $\la r \ra$.

The level-spacing distribution $P(s)$ is defined such that $P(s)\Delta s$ gives the probability of finding $s_i$ in the interval $[s, s+\Delta s]$. For an integrable system, the level-spacing distribution follows Poisson statistics, $P(s) \propto e^{-s}$ \cite{Berry_1977}, and the corresponding average ratio is $\la r \ra_{\text{Poisson}} = 2\ln 2 - 1$. In contrast, for non-integrable systems (with time-reversal symmetry) the level-spacing statistics is described by the Gaussian Orthogonal Ensemble (GOE). Specifically, we expect the Wigner-Dyson distribution $P(s) \propto s e^{-s^2}$ and $\la r \ra_{\text{GOE}} = 4 - 2\sqrt{3}$ \cite{Bohigas_1984, mehta_2004, Huse_2007,Atas_2013}.

\begin{figure}[t]

\includegraphics[width=\columnwidth]{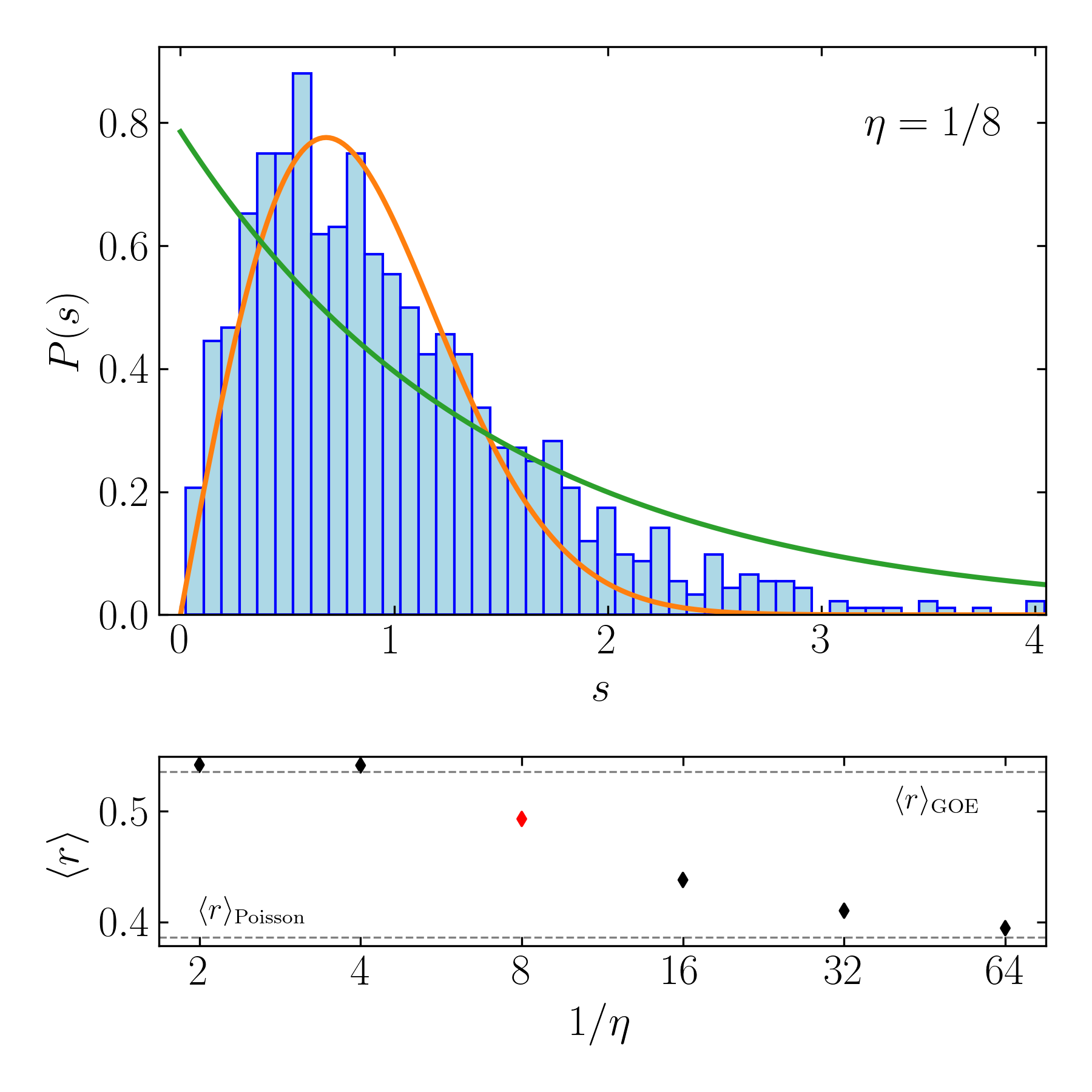}
\caption{
(a) Level-spacing distribution $P(s)$ of the model for $\eta = 1/8$. The curves show fits with the Poisson distribution (green) and with the Wigner-Dyson distribution expected from the GOE (orange).
(b) Average ratio $\langle r \rangle$ as a function of $1/\eta$. The red point highlights the value corresponding to $\eta = 1/8$.
Data are obtained from exact diagonalization with parameters $[J,L] = [1,16]$, in the zero-magnetization sector and within the $(\mathcal{I},\mathcal{F}) = (1,1)$ symmetry sectors.
}
 \label{fig:LSS}
\end{figure}

In Fig.~\ref{fig:LSS}, we show that for $\eta = 1/8$ (the experimentally relevant value) the model predominantly exhibits non-integrable behavior, as indicated by both the level-spacing distribution $P(s)$ and the average ratio $\la r \ra$. The spectrum is analyzed in the zero-magnetization sector, restricting to a fixed reflection-symmetry sector about the center of the chain, $\mathcal{I}: \sigma_{L/2+j}^{\alpha} \mapsto \sigma_{L/2-j}^{\alpha}$, and to a fixed spin-flip symmetry sector along the $z$ axis, $\mathcal{F}: (\sigma_j^x, \sigma_j^y, \sigma_j^z) \mapsto (\sigma_j^x, \sigma_j^y, -\sigma_j^z)$.

\paragraph{Details on the numerical simulations---}

In this section, we provide numerical details regarding the preparation of the initial state and its time evolution. The Fermi sea on a half chain at a given filling is prepared using the DMRG algorithm at fixed magnetization. The algorithm is stopped when the energy difference over five consecutive sweeps is below $10^{-10}$.
The resulting initial state is then evolved using a second-order TEBD algorithm~\cite{SCHOLLWOCK201196}. In particular, we use a time step $\delta t = 0.02$, a maximum bond dimension $\chi = 512$, and a truncation cutoff $\epsilon = 10^{-12}$, for which convergence of the results has been verified. 

\paragraph{Entanglement entropy---}

\begin{figure}[t]
\includegraphics[width=\columnwidth]{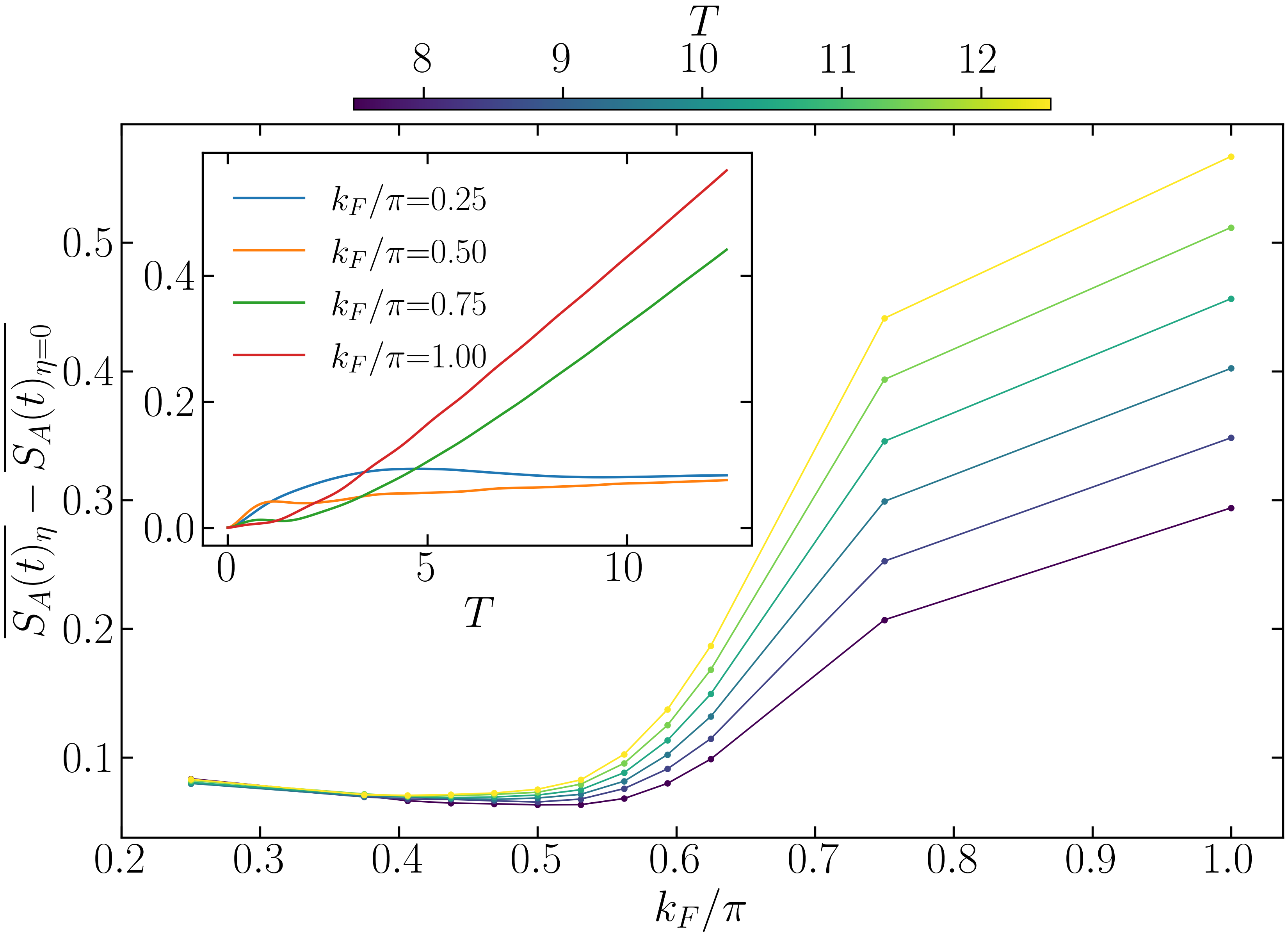}
 \caption{Difference between the half-chain entropy in the presence of the perturbation (at $\eta= 1/8$) and in its absence, averaged over the time window $t \in [0,T]$ and plotted as a function of $k_{\text{F}}$. An inset shows the same quantity as a function of $T$. The parameters are $[J,L] = [1, 64]$.}
 \label{fig:Transition_entangl}
\end{figure}

We study the growth of the von~Neumann entanglement entropy associated to half the system size, we denote it by $S_A$ (with $A = [-L/2,0]$), for different values of $k_{\text{F}}$.
As we did for the variance of the half-chain magnetization in Eq.~\eqref{eq:av_var},
we average it over the time windows $t \in [0,T]$ and we denote it by $\overline{S_A(t)}_\eta$. We plot the difference between the values at $\eta = 1/8$ and $\eta=0$ in Fig.~\ref{fig:Transition_entangl}. 
As in the main text, also here a qualitative change of behaviour is manifest. 

\paragraph{Higher-order effects---}

In the main text, we discuss the stability (or lack thereof) of the NESS in Eq.~\eqref{eq:n_NESS} from the perspective of two-body scattering. These processes are expected to be dominant as they arise directly from Fermi's golden rule, associated with second-order perturbation theory applied to Eq.~\eqref{eq:pert} where quartic fermionic terms are present. Conversely, if one considers higher-order effects or accounts for the additional neglected terms in the Hamiltonian, specifically those with $|j-i| \geq 3$ in Eq.~\eqref{eq:XY_ham}, genuine $n$-body scattering processes must be taken into account. In this section, we argue that although the latter are expected to manifest only at very large timescales, they inevitably destabilize the NESS \eqref{eq:n_NESS} for any finite value of $k_{\text{F}}$.

For instance, consider a process where $n$ particles with momenta $\{k_{j}\}_{j=1,\dots, n}$ (where $k_j \in (0,k_{\text{F}})$) undergo the scattering transition $k_{j} \rightarrow k'_{j} = -k_{j}$. The net variation of energy, defined as $\sum_{j=1}^{n} (\varepsilon(k'_j) -\varepsilon(k_j))$ vanishes identically in this configuration. Similarly, the total momentum transfer $\sum_j (k'_{j}-k_{j})$ vanishes modulo $2\pi$ whenever 
\begin{equation}\label{eq:condition_k}
\sum_{j=1}^{n} k_j = \pi.
\end{equation}
Note that since $k_{j} \leq k_{\text{F}}$, the condition in Eq.~\eqref{eq:condition_k} can only be satisfied if $k_{\text{F}} \geq \pi/n$. Because this process is kinematically allowed by both energy and momentum conservation, it is generically expected to occur. Consequently, for any given $k_{\text{F}} > 0$, one can always find a sufficiently large $n$ such that the corresponding $n$-body scattering destabilizes the NESS \eqref{eq:n_NESS}.
Since in our model only two-particle scattering is allowed, these processes can only arise as high-order effects in the interaction strength.
It is important to stress that the higher the order of the scattering process, the longer the timescale on which it is expected to affect the many-body dynamics.

\bibliography{bibliography}

\end{document}